\documentclass[twocolumn,showpacs,amssymb,nobibnotes,superscriptaddress,aps,prl]{revtex4-1}

\usepackage{enumerate,amsmath}
\usepackage{graphicx}

\def\mbr{\mathbf{r}}
\def\mbone{\mathbf{1}}
\def\mbtwo{\mathbf{2}}
\def\mbthree{\mathbf{3}}

\def\mba{\mathbf{a}}

\def\mbb{\mathbf{b}}

\def\mbZ{\mathcal{Z}}

\begin{document}

\title{Topologically protected measurement-based quantum computation
on the thermal state of a nearest-neighbor two-body Hamiltonian with spin-3/2 particles}%

\author{Keisuke Fujii}
\affiliation{
Graduate School of Engineering Science, Osaka University,
Toyonaka, Osaka 560-8531, Japan}

\author{Tomoyuki Morimae}
           \affiliation{
 Universit\'e Paris-Est Marne-la-Vall\'ee, 77454 Marne-la-Vall\'ee
 Cedex 2, France}
 \affiliation{
Interactive Research Center of Science (IRCS), Tokyo Institute of Technology,
2-12-1 Ookayama, Meguro-ku, Tokyo 152-8550, Japan
}
\date{\today}
\begin{abstract}
Recently, Li {\it et al.} [Phys. Rev. Lett. {\bf 107}, 060501 (2011)] have demonstrated 
that topologically protected measurement-based quantum computation can be implemented 
on the thermal state of a nearest-neighbor two-body Hamiltonian
with spin-2 and spin-3/2 particles
provided that the temperature is smaller than a
critical value, namely, threshold temperature.
Here we show that the thermal state of a nearest-neighbor two-body Hamiltonian,
which consists of only spin-3/2 particles, allows us
to perform topologically protected measurement-based quantum computation. 
The threshold temperature is calculated and
turns out to be comparable to that with the spin-2 and spin-3/2 system.
Furthermore, we generally show that a cluster state of 
high connectivity can be efficiently generated
from the thermal state of the spin-3/2 system
without severe thermal noise accumulation.
\end{abstract}


\maketitle

For the past two decades tremendous effort 
has been devoted to the realization of
quantum information processing 
both experimentally and theoretically.
It is, however, still under extensive investigation 
what type of physical system is best suited to the experimental realization 
of quantum information processing.
This question is closely related to the paradigm for quantum computation.
Measurement-based quantum computation (MBQC), 
which simulates the standard quantum circuit model
on the cluster states with adaptive measurements \cite{OWC},
significantly relaxes the requirements for experimental realization of quantum computation;
it allows us to perform scalable quantum computation even with probabilistic two-qubit gates 
 \cite{Nielsen04,Browne05,Duan05,Barrett05}.
Furthermore, topologically protected MBQC 
can be implemented on the three-dimensional (3D) cluster state
in a fault-tolerant way \cite{Raussendorf06,Raussendorf07,Li10,FT10}.

Despite these interesting achievements,
the cluster state cannot be the exact ground state 
of any naturally occurring Hamiltonian \cite{Nielsen06,Nest08}.
More generally, the ground state of any spin-1/2 
frustration-free Hamiltonian
with nearest-neighbor two-body interactions cannot 
be a universal resource for MBQC \cite{Chen11}.
This fact motivates us to seek MBQC on a higher dimensional system,
where the ground state is a universal resource \cite{Verstraete04}.
Recently a general framework, quantum computational tensor network (QCTN), 
for such MBQC on higher dimensional systems
has been developed \cite{Gross07},
where the resource states are represented by
matrix product states (MPSs) \cite{Fannes91} (or, more generally, tensor network states \cite{Verstraete08,Cirac09}).
Then, the one-dimensional Affleck-Kennedy-Lieb-Tasaki (AKLT) state with spin-1 particles 
has been shown to be universal with the help of dynamical coupling \cite{Gross07,Brennen08}.
Several genuine two-dimensional (2D) systems with spin-3/2 and spin-5/2 particles
have been discovered so far,
where the exact ground states of nearest-neighbor two-body Hamiltonians are universal resources for MBQC \cite{Chen09,Cai10,Wei11,Miyake10,Li11} (see Table \ref{table1}).
In real experimental setups, however,
it is more realistic
to assume that we can prepare the thermal equilibrium state
with a finite (possibly very low) temperature
instead of an exact ground state.
Furthermore,
measurements in MBQC themselves might also introduce imperfections.
To handle these issues,
fault-tolerant quantum computation is necessary.

\begin{table*}
\begin{center}
\begin{tabular}{c|c|c|c }
\hline \hline
dimension &  resource & model & threshold temperature
\\
\hline 
6 (spin-5/2) & ground state & 2D tri-cluster state by Chen {\it et al.} \cite{Chen09} &
\\
4 (spin-3/2) & ground state & quasi 1D AKLT by Cai {\it et al.} \cite{Cai10} &
\\
4 (spin-3/2) & ground state & 2D AKLT by  Wei {\it et al.} \cite{Wei11} and Miyake \cite{Miyake10} & 
\\
4 (spin-3/2) & ground state & 2D honeycomb by Li {\it et al.} \cite{Li11} &
\\
5 (spin-2) and 4 (spin-3/2) & thermal state & 3D lattice by Li {\it et al.} \cite{Li11} & $0.21 \Delta$ 
\\
4 (spin-3/2) & thermal state & 3D lattice in this paper & $0.18 \Delta $
\\
\hline \hline
\end{tabular}
\caption{Summary of the models where the ground state or the thermal state
of each two-body nearest-neighbor Hamiltonian can be used as a resource state
for universal MBQC with single-particle measurements.}
\label{table1}
\end{center}
\end{table*}%

Recently,
Li {\it et al.} \cite{Li11} have shown 
that the ground state of a certain nearest-neighbor two-body Hamiltonian 
of spin-3/2 particles in 2D can be used as a universal resource state 
for MBQC.
Furthermore, they have also shown that
the thermal state of a certain  nearest-neighbor two-body Hamiltonian of 
spin-2 center and spin-3/2 bond particles in 3D
can be used as a resource state
for topologically protected measurement-based quantum computation (TMBQC) \cite{Raussendorf06,Raussendorf07}
provided the temperature
is smaller than a certain critical value, namely {\it threshold temperature}.
It is, however, still unclear whether or not 
the thermal state of a nearest-neighbor two-body Hamiltonian
which consists of only spin-3/2 or spin-1 particles
can be useful for MBQC with single particle measurements.

In this Rapid Communication, 
we show that we can actually perform TMBQC
on the thermal state of a certain  nearest-neighbor two-body Hamiltonian 
which consists of only spin-3/2 particles.
Instead of the spin-2 center particle in Ref. \cite{Li11},
we utilize two spin-3/2 center particles jointed by one spin-3/2 bond particle.
In such a case, additional particles are required,
and therefore 
one might think that the threshold temperature is degraded significantly
due to the thermal noise accumulation.
However, it is not the case.
We calculate the threshold temperature of the TMBQC,
and it turns out to be comparable to 
that of the spin-2 and spin-3/2 system \cite{Li11}.
This is due to the fact that 
the thermal noise is suppressed exponentially at a low temperature
because of the energy gap,
and therefore the thermal noise accumulation can be compensated 
by decreasing the temperature slightly.
We further extend this result
to prepare a cluster state of high connectivity,
such as the star-cluster \cite{Li10,FT10}.
We generally show that
a cluster state in which each qubit is connected to at most $m$ other qubits,
say, the $m$-connected cluster state,
can be generated from 
the thermal state of a spin-3/2 system
by using only single particle measurements,
where, instead of the spin-$m/2$ center particle \cite{Li11},
$(m-2)$ spin-3/2 center particles are employed.
In such a case, more particles depending on the connectivity $m$ are required
compared to the method pointed out in Ref. \cite{Li11} with the spin-$m/2$ system.
However, we show that, 
by increasing the inverse temperature $\Delta \beta \rightarrow \Delta \beta + \ln(m-2)$,
an $m$-connected cluster state can be 
created from a thermal state of the spin-3/2 system
with a comparable fidelity to the $3$-connected cluster state
created from a thermal state with an inverse temperature $\Delta \beta$.
This result indicates that 
a cluster state of high connectivity can be 
prepared efficiently from the thermal state of a nearest-neighbor two-body Hamiltonian
with spin-3/2 particles by using only single-particle measurements.

Let us first review the spin-3/2 system 
introduced in Ref. \cite{Li11}.
The Hamiltonian is given by
\begin{eqnarray*}
 H&=& \Delta  \sum _{\mbr} \vec{S} _{\mbr} \cdot (\vec{I}_{ \mbr + \mbone}+\vec{I}_{\mbr +\mbtwo}+ \vec{I}_{\mbr +\mbthree} )
\end{eqnarray*}
where $\vec{S}_{\mbr} \equiv (S_{\mbr}^x,S_{\mbr}^y,S_{\mbr}^z)$ is the spin-3/2 operator of the center particle at the position $\mbr$ (see Fig. 1 in Ref. \cite{Li11}),
and $\vec{I}_{\mbr + \mba} = 
\vec A _{\mbr + \mba} $ or
$\vec B_{\mbr+ \mba}$
depending on the interaction types (line or dash),
where $\vec A _{\mbr + \mba}\equiv (A_{\mbr +\mba}^x,A_{\mbr +\mba}^y,A_{\mbr+\mba}^z)$ 
and $\vec B_{\mbr+ \mba}\equiv (B_{\mbr +\mba}^x,B_{\mbr +\mba}^y,B_{\mbr+\mba}^z)$
are two independent spin-1/2 operators
on the bond spin-3/2 (four-dimensional) particle at the position $\mbr + \mba$ 
($\mba = \mbone , \mbtwo, \mbthree$) (see Fig. 1 in Ref. \cite{Li11}).
The above Hamiltonian $H$ can be reformulated as
\begin{eqnarray*}
H
&=&  \sum _{\mbr} H_{\mbr} =\Delta /2 \sum _{\mbr}
( \vec{T}^2  _{\mbr} - \vec{S}^2 _{\mbr}  -  \vec {I}_{\mbr} ^2)
 \end{eqnarray*}
where $\vec{I}_{\mbr} \equiv 
 \vec{I} _{\mbr + \mbone}+\vec{I} _{\mbr + \mbtwo}+\vec{I} _{\mbr + \mbthree}$
 and $\vec T _{\mbr} \equiv \vec S_{\mbr} + \vec I _{\mbr}$.
The ground state $|G \rangle = \bigotimes _{\mbr} |g_{\mbr} \rangle$ 
is given by $T_{\mbr} =0$, $S_{\mbr} =3/2$ and $I_{\mbr} =3/2$,
where $L_{\mbr}(L_{\mbr}+1)$ ($L=T,S,I$) is the eigenvalue 
of the operator $\vec L_{\mbr}^2$.
Each center particle in the ground state $|G\rangle$ 
is filtered by using the positive operator valued measure (POVM) measurement \{ $F^{\alpha}= ({S_{\mbr} ^\alpha} ^2 - 1/4)/\sqrt{6}$ \} ($\alpha =x,y,z$).
If the measurement outcome is $\alpha =z$,
we obtain a four-qubit GHZ state 
as the post POVM measurement state: 
\begin{eqnarray*}
|{\rm GHZ} ^4_{\mbr} \rangle  \equiv \frac{1}{\sqrt{2}} (|\tilde 0 +++\rangle + |\tilde 1 ---\rangle) ,
\end{eqnarray*}
where $-|\tilde 1 \rangle  $ and 
$|\tilde 0 \rangle $ are eigenstates of $S^z$ with eigenvalues $+3/2$ and $-3/2$, respectively, and 
$|\pm\rangle$ are the eigenstates of $A^{z}$ or $B^{z}$ with eigenvalues $\pm 1$, respectively
\cite{comment}.
Even if we obtain other outcomes,
we can transform the post POVM measurement state to $|{\rm GHZ} ^4_{\mbr}\rangle$
by local operations.
The four-qubit GHZ state is subsequently used to 
construct the 2D honeycomb cluster state,
which is a universal resource for MBQC, 
by measuring the operators $A^z \otimes B^x$ and $A^x \otimes B^z$
on the bond particle.
Below, an operator $A \otimes B$ is denoted as $AB$ for simplicity.

\begin{figure}
\centering
\includegraphics[width=65mm]{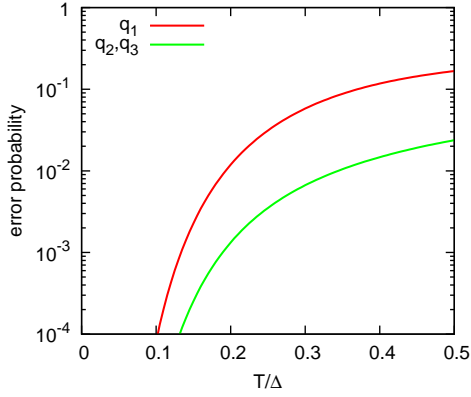}
\caption{(Color online). The error probabilities $q_{1}$ (red) and $q_2=q_3$ (green) are plotted as functions of the temperature $T/\Delta$.}
\label{errorprob}
\end{figure}

\begin{figure}[t]
\centering
\includegraphics[width=80mm]{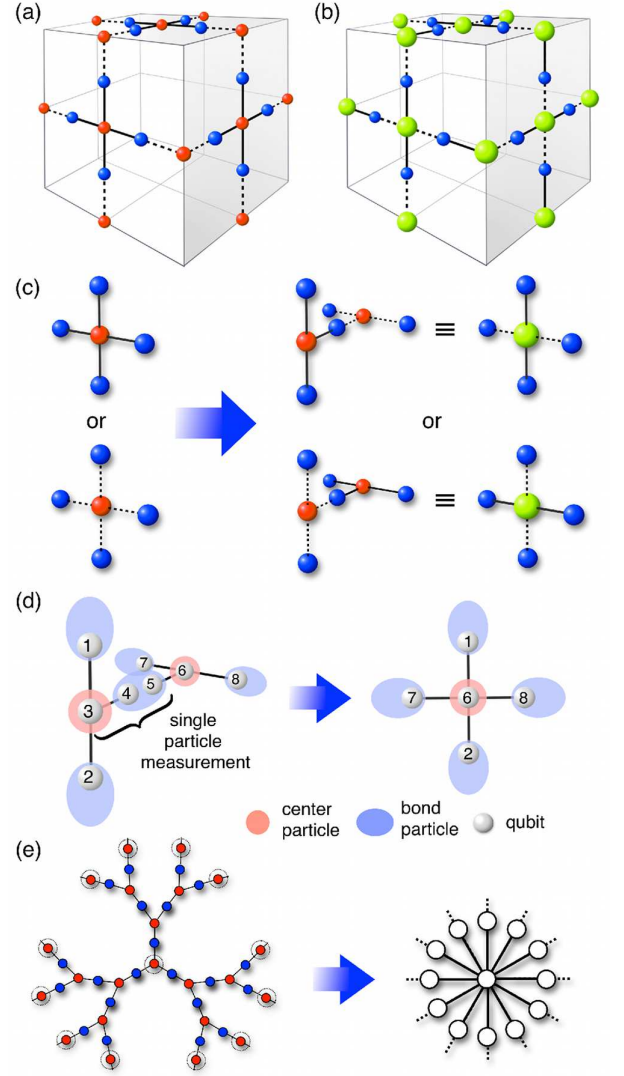}
\caption{(Color online). (a) The unit cell of the model by Li {\it et al.} \cite{Li11}. (b) The unit cell of the present model with only spin-3/2 particles. (c) In the present model, each center particle connected with four bond particles in the Li {\it et al.}'s model \cite{Li11} is replaced by two center particles, each of which is connected with three bond particles. (d) One of two center particles and the bond particle between them are 
measured locally to obtain the five-qubit GHZ state.
(e) An $m$-connected cluster state is created from 
a thermal state of spin-3/2 particles on the lattice of a tree structure
by single particle measurements. The center particles with dotted circles
form the $m$-connected cluster state. }
\label{Figure}
\end{figure}

In the case of a finite temperature, instead of the ground state,
we have the thermal state $\bigotimes _{\mbr} \rho _{\mbr}$
with $\rho _{\mbr} \equiv e^{ - \beta  H_{\mbr} } /\mbZ $,
where $\mbZ$ indicates the partition function and $\beta = T ^{-1}$
with a temperature $T$.  
Then, the GHZ state becomes a noisy one, say thermal GHZ state,
$\sigma _{\mbr} \equiv F^{\alpha } \rho _{\mbr}
{  F^{\alpha}} ^{\dag}/{\rm Tr}[ F^{\alpha } \rho _{\mbr}{  F^{\alpha}} ^{\dag}]$.
At a low temperature case,
the thermal GHZ state can be well approximated by $\mathcal{E}_4 (|{\rm GHZ} ^4_{\mbr} \rangle  \langle {\rm GHZ} ^4_{\mbr}| )$ with
\begin{eqnarray}
\mathcal{E}_4
&=& (1- q_1 - 3q_2 - 3q_3 ) [I]
+ q_1 [Z_{\mbr}  ]
\nonumber \\&&
+q_2 \sum _{\mba=\mathbf{1},\mathbf{2},\mathbf{3}} [Z_{\mbr + \mba}]
+q_3 \sum _{\mba=\mathbf{1},\mathbf{2},\mathbf{3}} [Z_{\mbr }Z_{\mbr + \mba}] ,
\label{error}
\end{eqnarray}
where $q_1$, $q_2$ and $q_3$ are error probabilities as functions of the temperature $T$,
$Z_{\mbb}$ is the Pauli $Z$ operator on the qubit at the position $\mbb$,
and $[C]\rho \equiv C \rho C^{\dag}$.
The probability of other errors such as $Z_{\mbr} Z_{\mbr +\mba} Z_{\mbr +\mba '}$
is several orders of magnitude smaller than $q_{1,2,3}$ \cite{comment2}.
In Fig. \ref{errorprob},
the error probabilities $q_{1,2,3}$ are plotted as functions of the temperature $T/\Delta$.
To obtain the 3D cluster state for TMBQC, in Ref. \cite{Li11}, 
the five-qubit GHZ state $|{\rm GHZ}^5_{\mbr}\rangle$ is generated similarly in the spin-2 and spin-3/2 system,
where center (red circle) and bond (blue circle) particles are spin-2 and spin-3/2, respectively,
as shown in Fig. \ref{Figure} (a).
A thermal version of the five-qubit GHZ state 
can be written similarly to Eq. (\ref{error})
by replacing $\sum _{\mba = \mathbf{1},\mathbf{2},\mathbf{3}}$
with $\sum _{\mba = \mathbf{1},\mathbf{2},\mathbf{3},\mathbf{4}}$.

Here we propose a different approach to prepare the 3D cluster state
for TMBQC.
Instead of the spin-2 center particle, 
which is connected to four spin-3/2 particles,
we use two spin-3/2 center particles,
each of which is connected to three spin-3/2 particles
as shown in Fig. \ref{Figure} (c).
By doing so, the unit cell of the present model is given as shown in Fig. \ref{Figure} (b),
where the replaced two center particles and one bond particle between them
are denoted symbolically by a green (light gray) circle .
The thermal state of such a system is given by $\bigotimes _{\mbr} \rho _{\mbr}$,
where $\mbr$ indicates the position of the replaced center particles.
After the filtering operation and local operations,
we have $\bigotimes _{\mbr} |{\rm GHZ} ^4_{\mbr} \rangle \langle {\rm GHZ} ^4_{\mbr}|$.
In order to obtain the 3D cluster state for TMBQC,
we first measure the observables $M^{(1)}=Y_{3}$, $M^{(2)}=Y_4 Z_5$, and $M^{(3)}=Z_4 Y_5$ 
on certain center and bond particles as shown in Fig. \ref{Figure} (d),
where $C_i$ ($C=X,Y,Z$) indicates the Pauli operator on the $i$th qubit.
Depending on the measurement outcomes $m_{i}$ 
of the operators $M^{(i)}$ ($i=1,2,3$),
the stabilizer operators for the post-measurement state are given by
\begin{eqnarray*}
\{
(-1)^{m_1+m_2} Z_1 Z_2  X_6 Z_7 Z_8, 
(-1)^{m_2+m_3} X_1Z_6,  
\\
(-1)^{m_2+m_3}X_2 Z_6, 
X_7 Z_6, X_8 Z_6 
\}.
\end{eqnarray*}
The above stabilizer operators are used to determine the error propagations
from the measured particles to the remaining particles
(e.g. the $Z_3$ error affects on the measurement outcome $m_1$ and 
appears as the $Z_6$ error on the post-measurement state).
By considering the above error propagations,
one can calculate the post-measurement
five-qubit GHZ state $\mathcal{E}_5 (|{\rm GHZ}_{\mbr} ^5 \rangle \langle {\rm GHZ}_{\mbr} ^5|)$ as follows:
\begin{eqnarray*}
\mathcal{E}_5 &=&
(1 - 2q_1 - 6q_2 - 6q_3) [I]
\\&&
+\sum _{i=1,2,7,8} (q_2 [Z_i]+q_3[Z_i Z_6])+ 2q_1 [Z_6]
+ 
\\&&
 (q_2+q_3)([Z_1 Z_2] + [Z_1 Z_2 Z_6]).
\end{eqnarray*}
We next measure the observable $A^x B^z$ and $A^z B^x$
on the remaining bond particles of the five-qubit GHZ states,
and finally the 3D cluster state for TMBQC is obtained.
Due to the measurements which connect five-qubit GHZ states into the 3D cluster state,
the $Z$ errors on the five-qubit GHZ state are propagated 
and finally located on the qubits in the 3D cluster state.
The probability of the independent errors on each qubit in the 3D cluster state
is given by $q_{\rm ind}=2q_1 +5q_2 +9q_3$.
The probability of the correlated errors, which are located on the qubits on each pair of opposite edges of each face in the 3D lattice \cite{Raussendorf06},
is given by $q_{\rm cor} =q_2 + q_3$.
(Note that the correlated errors between center and bond particles, e.g., $Z_1Z_6$, 
are corrected independently on the primal and dual lattices.
Thus they can be treated as independent errors \cite{Raussendorf06,Raussendorf07}.)
By using the threshold curve calculated in Ref. \cite{Raussendorf06},
the threshold temperature for the present model
is calculated to be $T=0.18\Delta$.
In the present approach,
more particles are required to generate the 3D cluster state for TMBQC
compared with Li {\it et al.}'s model,
and therefore one might think that the thermal noise is accumulated significantly.
However, the present threshold temperature $T=0.18\Delta$ with only spin-3/2 particles is 
comparable to that $T=0.21\Delta $ of the spin-2 and spin-3/2 system by Li {\it et al.} \cite{Li11}.
This is due to the fact that the thermal noise is suppressed 
exponentially at a low temperature because of the energy gap,
and hence one can suppress the error accumulations 
by slightly lowering the temperature.

This is also the case for the preparation of a general $m$-connected cluster state,
where $(m-2)$ spin-3/2 particles are employed as shown in Fig. \ref{Figure} (e).
Since the leading error of the four-qubit thermal GHZ state is 
$q_1 [Z_{\mbr}]$ as seen in Fig. \ref{errorprob},
the fidelity between $\mathcal{E}(|{\rm GHZ}^4 \rangle \langle {\rm GHZ}^4|)$
and $|{\rm GHZ}^4\rangle$ is given by $F_4( \Delta \beta) \simeq 1- q_1$,
where $q_1 \simeq 9e^{ - \Delta \beta}/5$ 
with a sufficiently small temperature ($ \Delta\beta \gg 1$).
On the other hand,
the leading error on the $m$-connected cluster state,
which is made from a thermal state of the spin-3/2 system,
where each $m$-connected qubit is realized by $(m-2)$ center particles,
is given by $(m-2)q_1[Z_{\mbr}]$ [where we assumed $(m-2) < 1/q_1$], and
the fidelity is calculated to be $F_m( \Delta\beta) \simeq 1- (m-2)q_1$.
Thus if the inverse temperature is increased to $ \Delta \beta '  \equiv  \Delta\beta +   \ln (m-2)$
one can achieve $F_m (\Delta \beta ' ) \simeq F_4 (\Delta \beta  )$.
This indicates that the thermal noise accumulation
can be efficiently suppressed by increasing the inverse temperature logarithmically
against the number of connection $m$ of the cluster state.

In conclusion,
we have shown that we can perform TMBQC 
on the thermal state of the nearest-neighbor two-body Hamiltonian
with only spin-3/2 particles.
It reduces the physical dimension of particles required to prepare the 3D cluster state for TMBQC
without significant degradation of the threshold temperature.
Similarly to the previous work by Li {\it et al.} \cite{Li11},
MBQC can be executed under always on interaction 
also in the present case. 
Furthermore, by extending the present approach,
we have also shown that an $m$-connected cluster state
can be efficiently generated from the thermal state of the spin-3/2 system,
where the thermal noise accumulation can be suppressed
by increasing the inverse temperature slightly.

Finally, we mention a possible route toward fully fault-tolerant
MBQC on the thermal state with a nearest-neighbor two-body Hamiltonian.
We (and Li {\it et al.} \cite{Li11}) have assumed that all operations, 
including filtering operations and single-particle measurements, 
on the thermal states are perfect so far.
In realistic experimental setups,
both filtering operations and single particle measurements
are subject to noise.
However, TMBQC could also deal with these imperfections.
More precisely, certain errors during these operations
lead to a leakage of the state from the computational basis (i.e., qubit).
If such a leakage error occurs
either in the filtering operation or in the single-particle measurement,
one can detect it, since, in such a case,
the outcomes of the filtering and measurement are inconsistent.
Thus such a detected leakage error can be corrected efficiently by using the loss-tolerant scheme \cite{Barrett}.
On the other hand, if errors occur in the filtering operation and measurement simultaneously,
they result in a map from the computational basis into itself.
Intuitively, such errors could be corrected as errors on the qubit during TMBQC, 
and hence they would not cause serious defects.
A detailed analysis of full fault-tolerance of MBQC in a higher dimension is an interesting topic 
for future work \cite{MF}.

The ground state (and hence the low-temperature thermal state) 
of any frustration-free nearest-neighbor two-body Hamiltonian
with spin-1/2 particles has been known to be useless for MBQC \cite{Chen11}.
We have shown here that
the thermal state of the nearest-neighbor two-body Hamiltonian with a spin-3/2 system
can be used as a resource state for MBQC, where 
the thermal noise can be corrected efficiently by TMBQC.
Now it is an interesting open question whether or not
we can perform TMBQC
on the thermal state of a nearest-neighbor two-body Hamiltonian which consists of spin-1 particles,
and, if it is possible, how high or low is the threshold temperature compared to that of the 
spin-3/2 system.

\begin{acknowledgments}
KF is supported by MEXT Grant-in-Aid for Scientific Research on Innovative Areas 20104003.
TM is supported by ANR (StatQuant JC07 07205763).
\end{acknowledgments}


\begin{thebibliography}{99}
\bibitem{OWC}
R. Raussendorf and H.-J. Briegel,
Phys. Rev. Lett. {\bf 86}, 5188 (2001);
R. Raussendorf, D. E. Browne, and H. J. Briegel,
Phys. Rev. A {\bf 68}, 022312 (2003).

\bibitem{Nielsen04}
M. A. Nielsen,
Phys. Rev. Lett. {\bf 93}, 040503 (2004).

\bibitem{Browne05}
D. E. Browne and T. Rudolph,
Phys. Rev. Lett. {\bf 95}, 010501 (2005).

\bibitem{Duan05}
L.-M. Duan and R. Raussendorf,
Phys. Rev. Lett. {\bf 95}, 080503 (2005).

\bibitem{Barrett05}
S. D. Barrett and P. Kok,
Phys. Rev. A {\bf 71}, 060310(R) (2005).

\bibitem{Raussendorf06}
R. Raussendorf, J. Harrington, and K. Goyal,
Ann. of Phys. {\bf 321}, 2242 (2006).

\bibitem{Raussendorf07}
R. Raussendorf and J. Harrington,
Phys. Rev. Lett. {\bf 98}, 190504 (2007);
R. Raussendorf, J. Harrington, and K. Goyal,
New J. Phys. {\bf 9}, 199 (2007).

\bibitem{Li10}
Y. Li, S. D. Barrett, T. M. Stace, and S. C. Benjamin,
Phys. Rev. Lett. {\bf 105}, 250502 (2010).

\bibitem{FT10}
K. Fujii and Y. Tokunaga,
Phys. Rev. Lett. {\bf 105}, 250503 (2010).


\bibitem{Nielsen06}
M. A. Nielsen,
Rep. Math. Phys.
{\bf 57}, 147 (2006).

\bibitem{Nest08}
M. Van den Nest, K. Luttmer, W. D{\" u}r, and H. J. Briegel,
Phys. Rev. A. {\bf 77}, 012301 (2008).

\bibitem{Chen11}
J. Chen, X. Chen, R. Duan, Z. Ji, and B. Zeng,
Phys. Rev. A {\bf 83}, 050301(R) (2011).


\bibitem{Verstraete04}
F. Verstraete and J. I. Cirac,
Phys. Rev. A {\bf 70}, 060302(R) (2004).

\bibitem{Gross07}
D. Gross and J. Eisert,
Phys. Rev. Lett. {\bf 98}, 220503 (2007);
D. Gross, J. Eisert, N. Schuch, and D. Perez-Garcia,
Phys. Rev. A {\bf 76}, 052315 (2007);
D. Gross, 
Ph.D. thesis, Imperial College London (2008).



\bibitem{Fannes91}
M. Fannes, B. Nachtergaele, and R. F. Werner,
J. Phys. A {\bf 24}, L185 (1991).

\bibitem{Verstraete08}
F. Verstraete, J. I. Cirac, and V. Murg, 
Adv. Phys. {\bf 57}, 143 (2008).

\bibitem{Cirac09}
J. I. Cirac and F. Verstraete, 
J. Phys. A: Math. Theor. {\bf 42}, 
504004 (2009).

\bibitem{Brennen08}
G. K. Brennen and A. Miyake,
Phys. Rev. Lett. {\bf 101}, 010502 (2008).



\bibitem{Chen09}
X. Chen, B. Zeng, Z.-C. Gu, B. Yoshida, and I. L. Chuang,
Phys. Rev. Lett. {\bf 102}, 220501 (2009).

\bibitem{Cai10}
J. Cai, A. Miyake, W. D{\" u}r, and H. J. Briegel,
Phys. Rev. A {\bf 82}, 052309 (2010).


\bibitem{Wei11}
T.-C. Wei, I. Affleck, and R. Raussendorf,
Phys. Rev. Lett. {\bf 106}, 070501 (2011).

\bibitem{Miyake10}
A. Miyake,
Ann. Phys. {\bf 326}, 1656 (2011).

\bibitem{Li11}
Y. Li, D. E. Browne, L. C. Kwek, R. Raussendorf, and T.-C. Wei,
Phys. Rev. Lett. {\bf 107}, 060501 (2011).





\bibitem{comment}
Note that in Ref. \cite{Li11}, the eigenstates of $A^{z}$ or $B^{z}$ with eigenvalues $\pm 1$
are defined as $|0\rangle$ and $|1\rangle$, respectively.
However, we here define the computational basis $|0\rangle$ and $|1\rangle$
so that $|\pm \rangle \equiv (|0\rangle \pm |1\rangle )/\sqrt{2}$ are the eigenstates
of $A^z$ or $B^z$ with eigenvalues $\pm 1$.
In this computational basis, the post POVM state with the filtering outcome $\alpha =z$
becomes a cluster state, which is equivalent to the GHZ state up to some Hadamard operations.
This simplifies the description of the thermal noise on the GHZ state.

\bibitem{comment2}
Note that errors on the GHZ state $|{\rm GHZ} ^4_{\mbr} \rangle$ can be rewritten 
as products of $Z$'s, since $|{\rm GHZ} ^4_{\mbr} \rangle$ is a cluster state in the present computational basis.

\bibitem{Barrett}
T. M. Stace, S. D. Barrett, and A. C. Doherty,
Phys. Rev. Lett. {\bf 102}, 200501 (2009);
S. D. Barrett and T. M. Stace, Phys. Rev. Lett. {\bf 105}, 200502 (2010).

\bibitem{MF}
T. Morimae and K. Fujii,
arXiv:1106.3720.
\end{thebibliography}
\end{document}